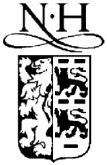

# Development of a time projection chamber with micro pixel electrodes

Hidetoshi Kubo[a,*], Kentaro Miuchi[a], Tsutomu Nagayoshi[a], Atsuhiko Ochi[b],
Reiko Orito[a], Atsushi Takada[a], Toru Tanimori[a], Masaru Ueno[a]

*[a]Department of Physics, Graduate School of Science, Kyoto University, Sakyo-ku, Kyoto 606-8502, Japan*

*[b]Department of Physics, Graduate School of Science and Technology, Kobe University, Kobe 657-8501, Japan*

## Abstract

A time projection chamber (TPC) based on a gaseous chamber with micro pixel electrodes (μ-PIC) has been developed for measuring three-dimensional tracks of charged particles. The μ-PIC with a detection area of 10 cm square consists of a double-sided printing circuit board. Anode pixels are formed with 0.4 mm pitch on strips aligned perpendicular to the cathode strips in order to obtain a two-dimensional position. In the TPC with drift length of 8 cm, 4 mm wide field cage electrodes are aligned at 1mm spaces and a uniform electric field of about 0.4 kV/cm is produced. For encoding of the three-dimensional position a synchronous readout system has been developed using Field Programmable Gate Arrays with 40 MHz clock. This system enables us to reconstruct the three-dimensional track of the particle at successive points like a cloud chamber even at high event rate. The drift velocity of electrons in the TPC was measured with the tracks of cosmic muons for three days, during which the TPC worked stably with the gas gain of 3000. With a radioisotope of gamma-ray source the three-dimensional track of a Compton scattered electron was taken successfully. © 2001 Elsevier Science. All rights reserved

*Keywords:* Gaseous detector; Time projection chamber; Micro-pattern detector

## 1. Introduction

The time projection chamber (TPC) is a three-dimensional tracking device for charged particles, which uses a two-dimensional array of pickup electrodes together with a measurement of the drift-time [1]. In a conventional TPC an avalanche occurs around the sense wire, and signals are induced on the cathode pads. However, a part of ions produced in the avalanche accumulate in the drift volume and distort the electric field. Furthermore the space charge increases in the avalanche region. In order to solve the problem, our Micro Pixel Chamber (μ-PIC) [2] classified as a micro-pattern detector [3] is proposed to provide signal electrodes with gas amplification. The development of a TPC based on the μ-PIC is reported in this paper. Other work on micro-pattern readout of TPCs is being done with GEMs [4] and Micromegas [5], as described in Ref. [6]. Our TPC could be of wide application in such fields as particle physics, medicine and astrophysics. Application to gamma-ray imaging is described in Ref. [7]. Fig. 1 shows the schematic structure of our μ-PIC. Around anode electrodes avalanche multiplication occurs due to a strong electric field, but the field is weaker at the edge of the cathode. Thus this structure should provide the higher gas gain than a MSGC without discharges [2, 8]. Since anode and cathode strips are arranged perpendicular to each other, two-dimensional readout is available. A μ-PIC with area of 10 cm square has been produced and its performance is reported in Ref. [9].

* Corresponding author. e-mail: kubo@cr.scphys.kyoto-u.ac.jp.

and each is connected to the next through 10 MΩ. Gas flows through the chamber via connectors at either side of the top plate. Fig. 2 also shows the calculated electric potential in the drift region, from Garfield7 [10]. An electric field of 0.4 kV/cm was produced uniformly in inner region.

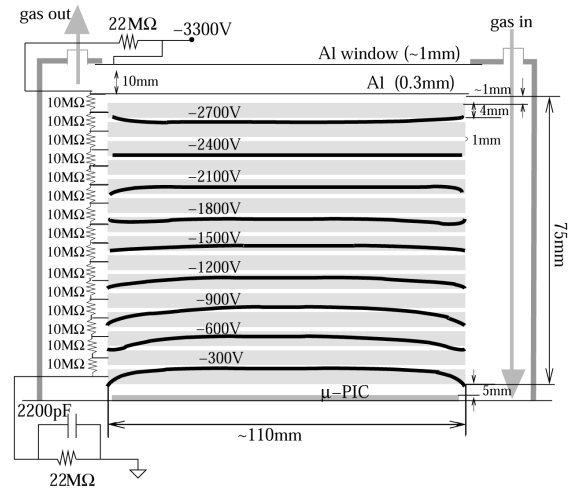

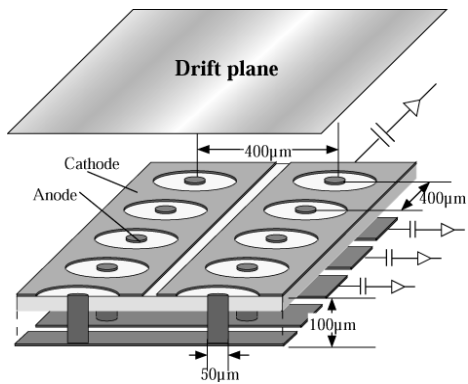

Fig. 1. Schematic structure of μ–PIC.

Fig. 2. Structure of the time projection chamber (upper panel) with the calculated electric potential (thick solid lines), and photograph (lower panel).

## 2. Design of TPC

Fig. 2 shows the structure of our TPC (hereafter micro-TPC) with area of 10 cm square and an 8 cm drift region. A negative high voltage of –3.3 kV is applied to the outer aluminum window. The 0.3 mm thick inner aluminum plane is connected to it through 22 MΩ. A field cage electrode of 4 mm width is connected to the plane through 10 MΩ. Another fourteen field cage electrodes are spaced by 1 mm,

## 3. Readout electronics system

A readout system for the micro-TPC has been developed, of which the block diagram is shown in Fig. 3. The μ-PIC is directly bonded onto a mother board with an area of 30 cm square. The anodes and cathodes are connected through 8 signal layers in the board to preamplifier cards attached on the rear side of the board. Each card has 8 fast ($\tau$ =16 ns) preamplifiers including discriminators with 4 input channels. These are ICs developed for ATLAS Thin Gap Chambers [11]. All discriminated signals in LVDS level from 256 anodes and 256 cathodes are fed to the position encoded system.

As described in Ref. [8], the fast pulses from both anodes and cathodes and synchronously encoding with a few 10 ns clock cycle enable us to encode more than $10^7$ events/s. For an event which generates more than three hit strips both on anodes and cathodes, a simple method of getting the hit position as a center of gravity of the hit electrodes can provide a position resolution of less than 200 μm. Thus a synchronous encoding system has been developed based on the system designed for MSGC [8]. The position encoding module (PEM) consists of 10 layers with area of 30 cm square and five Field Programmable Gate Arrays (FPGAs; Xilinx Virtex-E) accepting LVDS signals without any external terminator. This design reduces the size and the power consumption. The PEM accepts 512 outputs from discriminators, and then encodes hit electrodes to *X* or *Y* coordinates at 40 MHz clock. The information of the encoded position, event number, and clock counter are sent in LVDS level through 33 channel cables to a VME6U-based 32 bits memory module (MM), which has SRAM memories of 32 Mbyte in total. One of 33 channels is used for sending the 40 MHz clock generated in the PEM to the MM. To analyze the pulse shape from cathodes, the output signals of preamplifiers connected to the cathode electrodes are summed in units of 16 channels and fed to a VME-based 100 MHz flash

ADC (REPIC RPV-160). Then the data stored in the ADC and MM are read by a VME-based CPU (FORCE 7V) running the Solaris OS.

## 4. Performance of the micro-TPC

### 4.1. Position encoding

In order to check the position encoding, a TPC with a 1.2 cm drift region was operated in a Kr (90%) ethane (10%) mixture at 1 atm. The radioisotope $^{85}$Kr was present at 1 part in $10^{11}$. Two-dimensional images were taken by the position encoding system, as shown in Fig. 4. There were many tracks of electrons with 0.7 MeV emitted from $^{85}$Kr β-decay. The tracks consisted of successive points like a cloud chamber image.

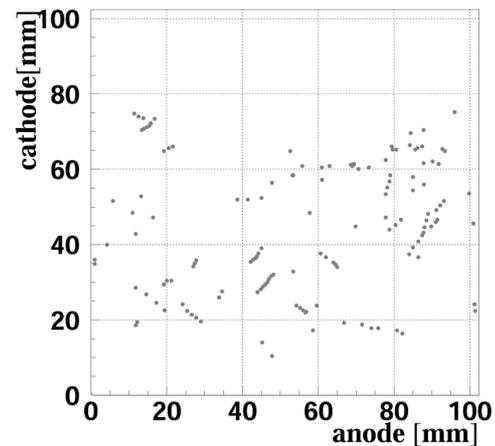

Fig. 4. Two-dimensional image taken with the μ-PIC, including β-decay electron tracks from $^{85}$Kr.

### 4.2. Gas gain and drift velocity

To measure the gas gain and the drift velocity of electrons in the TPC, using the tracks of cosmic muons, two plastic scintillators were placed at the both sides of the TPC perpendicular to the direction of the electric field. The coincidence of triggers from both scintillators was required. The gas mixture was argon 80% and ethane 20% at 1 atm. Fig. 5 shows the shapes of signals from cathode electrodes for one cosmic muon. The gas gain was estimated to be 3000.

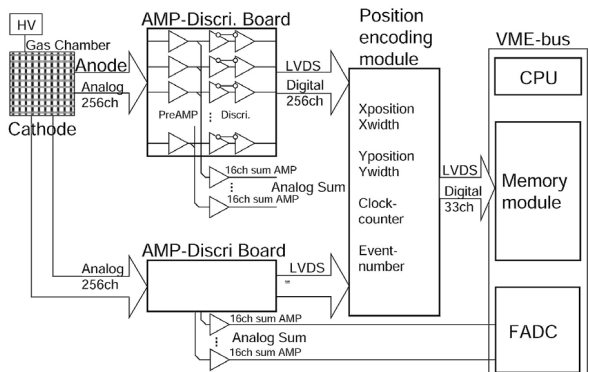

Fig. 3. Block diagram of the readout electronics system.

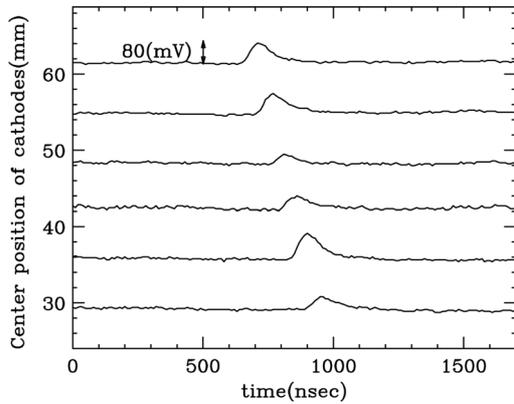

Fig. 5. Signal shapes from cathode electrodes for one incident cosmic muon event.

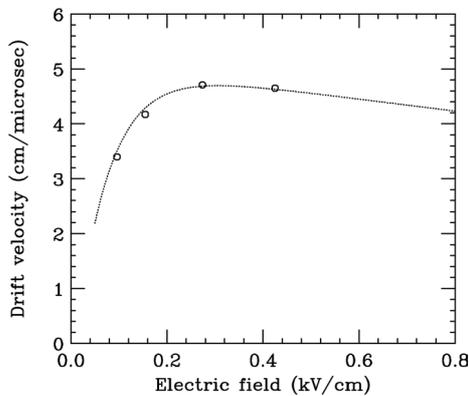

Fig. 6. Drift velocity of electrons in argon:ethane 80:20. The points are in this work, the dotted line from Ref. [12].

The rise time of the pulses suggests that a position resolution of 0.2 mm (rms) could be achieved with a suitable reconstruction algorithm. The detected count rate was constant for three days, and the TPC worked without deterioration [9]. The measured drift velocity of electrons, shown in Fig. 6, was consistent with a previous measurement [12].

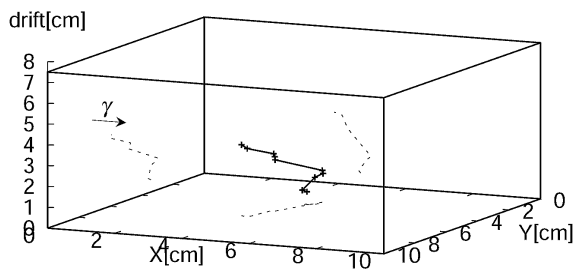

Fig. 7. Track of a Compton-scattered electron (solid line) and its projection on the planes (dot lines).

### 4.3. Three-dimensional track

So far the signal gain of the μ-PIC has not been high enough to measure three-dimensional tracks of cosmic muons. However the three-dimensional track of a Compton scattered electron has been reconstructed (Fig. 7), produced by a 511 keV annihilation gamma-ray from $^{22}$Na. The improvement of the μ-PIC is in progress for higher gas gain.

### 5. Summary

A time projection chamber (micro-TPC) using micro pixel electrodes for the signal readout with gas amplification has been developed. The TPC with the fast synchronous position encoding system enables us to perform the three-dimensional localization of a charged particle. The TPC worked stably for three days, and three-dimensional tracks of electrons were taken like a cloud chamber. However, the signal gain was rather low for a minimum ionizing particle. After the improvement of the μ-PIC, the TPC will be a promising tracking device with wide application.

### Acknowledgments

This work is supported by a Grant-in-Aid in Scientific Research of the Japan Ministry of Education, Science, Sports and Technology, and "Ground Research Announcement for Space Utilization" promoted by Japan Space Forum.